\documentclass[a4paper,11pt]{article}
\usepackage[T1]{fontenc}
\usepackage[utf8x]{inputenc}
\usepackage{color}
\usepackage{scalerel, soul}
\usepackage{bm}
\usepackage{stmaryrd, mathrsfs, dsfont, amsmath, latexsym}
\usepackage{youngtab, authblk}
\usepackage{hyperref}
\usepackage{verbatim}
\usepackage{cite}
\allowdisplaybreaks

\numberwithin{equation}{section}

\def \de{\partial}

\def \H{{\cal H}}

\def \N{{\cal N}}

\def \Tr{{\rm Tr}}
\def \Dof{{\rm Dof}}

\begin{document}

\title{One-loop quantum gravity from the ${\cal N}=4$ spinning particle}
\author[a,b,c]{Fiorenzo Bastianelli\, }
\author[d]{Roberto Bonezzi\, }
\author[e,b,c]{Olindo Corradini\, }
\author[f,b]{Emanuele Latini\, }

\affil[a]{\small\it  Dipartimento di Fisica e Astronomia, Universit\`a di Bologna, Via Irnerio 46, I-40126 Bologna, Italy}

\affil[b] {{\small\it INFN, Sezione di Bologna,  Via Irnerio 46, I-40126 Bologna, Italy}}

\affil[c]{{\small\it Max-Planck-Institut f\"ur Gravitationphysik, Albert-Einstein-Institut, Am M\"uhlenberg 1, 14476 Golm, Germany}}

\affil[d]{\small\it Institute for Physics, Humboldt University Berlin, Zum Gro{\ss}en Windkanal 6, D-12489 Berlin, Germany}

\affil[e]{{\small\it Dipartimento di Scienze Fisiche, Informatiche e Matematiche, \protect\\
Universit\`a degli Studi di Modena e Reggio Emilia, %\protect\\[-2.5mm] 
Via Campi 213/A, I-41125 Modena, Italy}}

\affil[f]{\small\it Dipartimento di Matematica, Universit\`a di Bologna, Piazza di Porta S. Donato 5, I-40126 Bologna, Italy}

\maketitle

\abstract{We construct a spinning particle that reproduces the propagation of the graviton on those curved backgrounds which 
solve the Einstein equations, with or without cosmological constant, i.e. Einstein manifolds. It is obtained by modifying the $\mathcal{N}=4$ supersymmetric 
spinning particle by relaxing the gauging of the full $SO(4)$ $R$-symmetry group to a parabolic subgroup, and selecting suitable 
Chern-Simons couplings on the worldline. We test it by computing the correct one-loop divergencies of quantum gravity in $D=4$.}

\section{Introduction}

%Quantum gravity has undoubtedly been a topic of paramount importance in theoretical physics for the past decades. However, to date, no completely satisfactory way of quantizing gravity has been achieved. The approach we take in the present manuscript involves the worldline path integral quantization of spinning particle models.

 Quantum gravity has undoubtedly been a topic of major interest in theoretical physics.
 Here we wish to study the graviton using a worldline approach. Some benefits of a first-quantized description are well known in the history of string theory: some properties, like the relation between gauge and gravitational amplitudes, T-duality and so on are quite transparent from the worldsheet description of the string, but pretty much hidden in terms of the target space effective field theory. Similar considerations apply to standard quantum field theories as well: Despite the great success of covariant perturbation theory for electrodynamics, the use of lagrangian Feynman rules to compute even simple QCD processes becomes soon intractable. Moreover, the relative simplicity of tree-level gluon scattering amplitudes, and their relation to graviton amplitudes (most naively stated as ``gravity is Yang-Mills squared'') are completely obscure in the standard lagrangian treatment of Yang-Mills and gravity. Therefore, our aim in the present paper is to continue developing a w!
 orldline description of perturbative quantum gravity, that is able to capture these features in a more transparent way. For instance, the worldline field content of the $\N=4$ spinning particle, that is the relevant model for gravity, consists of two copies (in the fermionic sector) of the worldline variables of the $\N=2$ model, relevant for Yang-Mills. Similarly, the graviton three-point function was easily shown  to exhibit the double copy structure as compared to Yang-Mills~\cite{Bonezzi:2018box}.  
 
However, the search of a worldline description for the graviton has met several obstructions along the years. Free spinning particle models based 
on worldline supersymmetry were suggested and constructed in \cite{Berezin:1976eg,Gershun:1979fb,Howe:1988ft}, 
and contained the $\mathcal{N}=4$  supersymmetric spinning particle that describes a massless point particle of spin 2 in $D=4$ flat spacetime dimensions.
These models, which are based on $O(N)$-extended worldline supersymmetry, were found to enjoy conformal invariance \cite{Siegel:1988ru, Siegel:1988gd}, but their coupling to curved backgrounds 
seemed to invalidate the first class algebra that is at the basis of the models. Couplings to (A)dS \cite{Kuzenko:1995mg}  
and conformally flat spaces \cite{Bastianelli:2008nm, Corradini:2010ia} were later found to be viable, but the problem remained how to use them as worldline models for 
describing perturbative quantum gravity. A first attempt to construct a worldline representation of the graviton was made in \cite{Bastianelli:2013tsa}.
Then, a much more elegant approach has been proposed recently in \cite{Bonezzi:2018box}, building on the BRST construction that  
proved to be successful in the description of particles of spin 1 with the $\mathcal{N}=2$ supersymmetric spinning model \cite{Dai:2008bh}.

The definition of the path integral on the circle (that is related to QFT 1PI one-loop amplitudes) poses, in particular, some conceptual issues. The main step forward in constructing consistent worldline descriptions of Yang-Mills \cite{Dai:2008bh} and gravity \cite{Bonezzi:2018box} was to realize that the corresponding BRST systems are consistent \emph{only} upon a suitable truncation of the Hilbert space. In the case of the $\N=2$ spinning particle this has an immediate field theoretic explanation, as $p$-forms admit Yang-Mills interactions only for $p=1\,$. Similarly, the minimal constraint in the $\N=4$ case is $U(1)\times U(1)\,$, that leaves the massless NS-NS spectrum of closed string theory, and fits with the field theory result of \cite{Bekaert:2004dz}. While this is not a problem for the BRST cohomology itself, nor for tree level amplitudes, one has to find the correct way to implement the projection at one-loop level, since in principle all the unwanted states pro!
 pagate in the loop. Moreover, the path integral on the circle is usually derived from gauge fixing a worldline action with local (super)symmetries, based on a first class constraint algebra. In the presence of a non-trivial gravitational background, the constraint algebra is obstructed and not first class anymore. A more appropriate way of thinking about the model is thus to consider it as a genuine BRST system from the start, regardless of being derived from a gauge invariant classical predecessor.
Our main goal here will be to determine the measure on the moduli space implementing the correct projection on the pure gravity contribution, thereby defining the path integral on the circle.

Thus, we study again the $\mathcal{N}=4$  spinning particle from this new prospective. First we consider the version defined by gauging the whole extended supersymmetry algebra of the worldline. In particular, we analyze the path integral on the loop (a worldline with the topology of a circle),
 constructed in \cite{Bastianelli:2007pv}, and dissect it to see how the gauge symmetries project the full Hilbert space 
to the one of the spin 2 particle, which remains as the only propagating degree of freedom.
Studying the role played by the measure on the moduli space, left over by the gauge-fixing, allows us to modify the measure to achieve an improved 
projection. The latter has the virtue of working in any spacetime dimensions, allowing also more general couplings to curved backgrounds.
This modification of the measure on moduli space is interpreted as due to the gauging a parabolic subgroup of the 
$SO(4)$ $R$-symmetry group, supplemented by appropriate Chern-Simons couplings on the worldline.

The model is eventually tested on curved backgrounds, were it is found to reproduce the correct results for the diverging part of the graviton one-loop 
effective action \cite{tHooft:1974toh, Christensen:1979iy}.

\section{Anatomy of degrees of freedom}

We are going to review the quantization on the circle of the free ${\N=4}$ spinning particle with various gaugings of the $SO(4)$ $R$-symmetry algebra. In particular, we focus on the precise way the path integral extracts the physical degrees of freedom from the Hilbert space.
This serves as a guiding principle for the quantization of the corresponding non-linear sigma model which couples a spin 2 particle (the graviton) 
to background gravity, and we use it to compute terms in the graviton one-loop effective action in arbitrary dimension, checking that it gives
the expected gauge invariant (BRST invariant)
result in four dimensions.

\subsection{Warm up: ${\N=2}$ and  $p$-forms}

Before analyzing the case of the ${\N=4}$ spinning particle, relevant for gravity, we shall review the quantization on the circle of the ${\N=2}$ particle, describing differential forms \cite{Howe:1989vn,Bastianelli:2005vk}, in order to display how the gauging of worldline supersymmetries extracts the physical degrees of freedom in a covariant way, that is related to the BV treatment of gauge theories in target space. This is maybe not surprising, since the BRST wavefunction of the particle contains the full BV spacetime spectrum, and the first-quantized BRST operator serves as BRST-BV differential in target space~\cite{Siegel:1989ip,Barnich:2003wj,Barnich:2004cr,Barnich:2006pc}. 

The action for the free ${\N=2}$ spinning particle in phase space reads
\begin{equation}
S=\int dt\,\big[p_\mu\dot x^\mu+i\,\bar\psi^\mu\dot\psi_\mu-e\,p^2-i\chi\,\bar\psi\cdot p-i\bar\chi\,\psi\cdot p-a(\psi\cdot\bar\psi-q)\big]  \;,  
\end{equation}
where the one dimensional supergravity multiplet $(e,\chi, \bar\chi, a)$ gauges worldline reparametrizations, supersymmetries and $U(1)$ $R$-symmetry, respectively. Here ``$\cdot$'' means contraction over spacetime indices $\mu, \nu, ...$ while  $x^\mu(t)$ and $p_\mu(t)$ are spacetime phase space variables, and $(\psi^\mu(t), \bar\psi^\mu(t))$ are worldline fermions, analogous to the RNS fermions of the spinning string. In order to project on $p$-form gauge fields we have included a Chern-Simons coupling\footnote{We use a symmetric (Weyl) ordering of the quantum operators, e.g. $\psi\cdot\bar \psi = \frac12 (\psi\cdot\bar\psi-\bar\psi\cdot\psi)\to\hat N_\psi-\frac D2$, where $\hat N_\psi=\psi\cdot\frac{\de}{\de\psi}$ is the $\psi-$number operator. This matches with the path integral regularization we use.} $q=p+1-\tfrac{D}{2}$ that converts the classical constraint $(\psi\cdot\bar\psi-q)$ into the operator $(\hat N_\psi-p-1)\,$ \cite{Bastianelli:2005vk}. The contribution fro!
 m the ghosts will then shift the eigenvalue to $p$ of the $p$-form gauge field.

The euclidean action in configuration space becomes
\begin{equation}\label{euclidaction}
S=\int d\tau\,\big[\tfrac{1}{4e}\,(\dot x^\mu-\chi\bar\psi^\mu-\bar\chi\psi^\mu)^2+\bar\psi^\mu(\de_\tau+ia)\psi_\mu+iq\,a\big]  \;,  
\end{equation}
upon Wick-rotating the gauge field $a\to-ia$ to maintain the $U(1)$ group compact. 
On the circle we gauge fix the supergravity multiplet as $(e,\chi,\bar\chi,a)\to(T,0,0,\theta)$ and the nontrivial ghost action is given by
\begin{equation}
S_{\rm gh}=\int d\tau\,\big[\bar\beta(\de_\tau+i\theta)\gamma+\beta(\de_\tau-i\theta)\bar\gamma\big] \;,    
\end{equation}
for the system of bosonic superghosts.
The partition function (that multiplied by $-\frac12$
corresponds to the QFT effective action) is then given by
\begin{equation}
Z_p=\int_0^\infty\frac{dT}{T}\,Z_p(T)\;,\quad Z_p(T):=\int_0^{2\pi}\frac{d\theta}{2\pi}\int_{\rm P}Dx\int_{\rm A}D\bar\psi D\psi\,\int_{\rm A} D\bar\beta D\gamma D\beta D\bar\gamma\, e^{-S_{\rm gf}-S_{\rm gh}}  \;,
\end{equation}
where $S_{\rm gf}$ denotes the gauge fixed version of \eqref{euclidaction}, and the subscripts on the functional integrals denote (anti)-periodic boundary conditions.
In order to unveil the spacetime gauge structure, it is instructive to recast the density $Z_p(T)$ in operator terms as
\begin{equation}
Z_p(T)=\int_0^{2\pi}\frac{d\theta}{2\pi}\,\left(2\cos\tfrac{\theta}{2}\right)^{-2}\Tr\left[e^{-T\hat H}e^{i\theta(\hat N_\psi-p-1)}\right]\;,
\end{equation}
where $\hat H=\hat p^2$ for the free theory, and the trace is over the Hilbert space consisting of differential forms of arbitrary degree contained in the $\psi$ Taylor coefficients of wavefunctions $\omega(x,\psi)\,$. The factor of $\left(2\cos\tfrac{\theta}{2}\right)^{-2}$ comes from the path integral over the SUSY ghosts, and is responsible for the spacetime gauge structure, as we will briefly review.
To proceed further, let us notice that the Hilbert space can be decomposed as a direct sum according to the form degree, \emph{i.e.} the eigenvalues of $\hat N_\psi$, as $\H=\bigoplus_{n=0}^D\H_n\,$. Accordingly, the trace can be decomposed as well and we shall denote
\begin{equation}
t_n(T):=\Tr_n\left[e^{-T\hat H}\right]  \;.  
\end{equation}
For the free particle one has $t_n(T)=\binom{D}{n}\,\tfrac{1}{(4\pi T)^{D/2}}\,$, where the $T$-factor corresponds to the free particle position, while the binomial simply gives the corresponding number of independent components of an antisymmetric tensor of rank $n\,$. One can now use a Wilson line variable $z:=e^{i\theta}$ and find
\begin{equation}
Z_p(T)=\oint_{\gamma^-}\frac{dz}{2\pi iz}\,\frac{1}{(z+1)^2}\sum_{n=0}^Dt_n(T)\,z^{n-p}  = \sum_{k=0}^p(-)^k(k+1)\,t_{p-k}(T)  \;,
\end{equation}
where we have slightly deformed the contour $\lvert z\rvert=1$ to exclude the pole in $z=-1\,$. The above result coincides with the decomposition one would obtain from the BV action in field theory: the contribution for $k=0$ being the gauge $p$-form, $k=1$ the $(p-1)$-form ghost, $k=2$ the first ghost for ghost and so on. It is clear, from the way the expansion is extracted from the above formula, that $\frac{z}{(z+1)^2}$ is the crucial factor for obtaining the correct residues and, as anticipated, it comes from the superghosts\footnote{This should be expected, since in the light-cone formulation the local supersymmetries explicitly remove the unphysical polarizations.}. In order to treat the SUSY ghost sector in a similar footing to the $\psi$ ``matter'' sector, we now rewrite the same density by undoing the path integral over the $\beta\gamma$-system:
\begin{equation}\label{N=2BRST}
Z_p(T)=\int_0^{2\pi}\frac{d\theta}{2\pi}\,\Tr_{\rm BRST}\left[e^{-T\hat H}e^{i\theta(\hat\N-p)}(-)^{\hat N_\gamma+\hat N_\beta}\right]= \Tr_{\rm BRST}\left[e^{-T\hat H}\delta(\hat\N-p)(-)^{\hat N_\gamma+\hat N_\beta}\right]\;, 
\end{equation}
where the symbol $\delta(\hat\N-p)$ is a Kronecker delta, that selects the contribution of sectors with $\hat \N$-occupation number 
equals to $p$.   
Here $\hat\N:= \psi^\mu \bar\psi_\mu + \gamma \bar\beta - \beta \bar\gamma $ and we observe that on the BRST vacuum annihilated by $\bar\gamma$ and $\bar\beta$ it also takes the form  ${\hat \N} = N_\psi + N_\gamma + N_\beta$; the trace is over the BRST Hilbert space\footnote{This coincides with the treatment given in \cite{Dai:2008bh} for spin one.} with vacuum annihilated by $\bar\beta$ and $\bar\gamma\,$. We shall notice that the alternating signs appearing in \eqref{N=2BRST} are due to the antiperiodic boundary conditions in a bosonic path integral, and have the spacetime interpretation of assigning negative contributions to the effective action (and so to the degrees of freedom) to fields of odd ghost number. 
To exemplify the above, let us consider the case of $p=2\,$: The BRST wavefunction at $\hat\N=2$ is given by\footnote{Prior to integrating the $(b,c)$ reparametrization ghosts, the BRST Hilbert space is doubled by the presence of $c$ in the wavefunction. This corresponds to the full BV spectrum in spacetime, with all the antifields included. Here, having already integrated out the $(b,c)$ system, the wavefunction corresponds to taking Siegel's gauge $b\,\Psi=0\,$.}
\begin{equation}
B_{\mu\nu}\,\psi^\mu\psi^\nu+\lambda_\mu\,\psi^\mu\beta+\lambda_\mu^*\,\psi^\mu\gamma+\phi\,\gamma\beta+\lambda\,\beta^2+\lambda^*\,\gamma^2    
\end{equation}
where $B_{\mu\nu}$ is the two-form gauge field, $(\lambda_\mu,\lambda_\mu^*)$ are the ghost-antighost vectors, $\phi$ is an auxiliary scalar and $(\lambda,\lambda^*)$ is the ghost for ghost paired with its canonical conjugate, giving immediately 
\begin{equation}
Z_2(T)=t_2(T)-2t_1(T)+3t_0(T)=\tfrac{1}{(4\pi T)^{D/2}}\, \tfrac{(D-2)(D-3)}{2} \;,  
\end{equation}
that corresponds to the physical transverse degrees of freedom of a two-form.

\subsection{${\N=4}\,$: gravitons and NS-NS spectrum}

We turn now to the same analysis for the case of the ${\N=4}$ spinning particle, relevant for gravity, with various gaugings of the $R$-symmetry group $SO(4)\,$. The corresponding phase space action reads
\begin{equation}
S=\int dt\,\big[p_\mu\dot x^\mu+i\,\bar\psi^{i\,\mu}\dot\psi_{i\,\mu}-e\,p^2-i\chi_i\,\bar\psi^i\cdot p-i\bar\chi^i\,\psi_i\cdot p-a^r\,J_r\big]  \;,  
\label{eq:4flat}
\end{equation}
where $i=1,2$ is a $U(2)$ index and $J_r$ denotes the subset of $SO(4)$ generators being gauged---we keep manifest only the symmetry $U(2)\subset SO(4)$. The states in the Hilbert space correspond to bi-forms,
interpreted as gauge fields:
\begin{equation}\label{wavefunction}
\omega(x,\psi_i)=\sum_{m,n=0}^D\omega_{\mu[m]\vert \nu[n]}(x)\,\psi_1^{\mu_1}...\psi_1^{\mu_m}\,\psi_2^{\nu_1}...\psi_2^{\nu_n}\sim\bigoplus_{m,n}\;m\left\{\,\yng(1,1,1,1)\right.\,\otimes\, n\left\{\yng(1,1,1)\,\right.\;.
\end{equation}
Being interested in gravity we will always project the above spectrum to the subspace $m=n=1$ for the graviton to be identified with the symmetric and traceless component of $\omega_{\mu|\nu}\,$. In this respect, the most economic choice is to gauge the $U(1)\times U(1)$ subgroup generated by
\begin{equation}
\hat N_i := \hat\psi_i\cdot\hat{\bar\psi}^i\;,\quad \text{index $i$ not summed}   \;, 
\end{equation}
where we used hats to stress that the above expression refers to operators in that given order. In this case one can add two independent Chern-Simons couplings $q_i\,$, that we choose as $q_i=2-\tfrac{D}{2}$ in order to project on the gravity sector. The spectrum consists of a graviton, a two-form and a scalar, coinciding with the massless NS-NS sector of closed strings. On the circle, the two gauge fields $a_i$ give rise to the angular moduli $(\theta, \phi)$ and, similarly to the ${\N=2}$ case, we obtain
\begin{equation}
Z_{U(1)\times U(1)}(T)=\int_0^{2\pi}\frac{d\theta}{2\pi}\int_0^{2\pi}\frac{d\phi}{2\pi}\,\left(2\cos\tfrac{\theta}{2}\right)^{-2}\left(2\cos\tfrac{\phi}{2}\right)^{-2}\Tr\left[e^{-T\hat H}e^{i\theta(\hat N_1-2)+i\phi(\hat N_2-2)}\right]\;.
\end{equation}
The Hilbert space, and so the trace, has an obvious double grading according to the eigenvalues of $\hat N_i\,$. We will thus define 
\begin{equation}
t_{m,n}(T):=\Tr_{m,n}\left[e^{-T\hat H}\right]  \;,
\end{equation}
allowing to write
\begin{equation}
Z_{U(1)\times U(1)}(T)=\oint_{\gamma^-}\frac{dz}{2\pi iz}\oint_{\gamma^-}\frac{dw}{2\pi iw}\,\frac{z}{(z+1)^2}\frac{w}{(w+1)^2}\sum_{n,m=0}^Dt_{n,m}(T)\,z^{n-2}w^{m-2} \;, 
\end{equation}
where the Wilson line variables are given by $z:=e^{i\theta}$ and $w:=e^{i\phi}\,$. The contour integrals are easily performed, yielding\footnote{For brevity we omit the dependence on $T$.}
\begin{equation}
Z_{U(1)\times U(1)}=t_{1,1}-2\,t_{1,0}-2\,t_{0,1}+4\,t_{0,0}    \;.
\end{equation}
In the free theory one has $t_{n,m}(T)=t_{m,n}(T)=\tfrac{1}{(4\pi T)^{D/2}}\binom{D}{n}\binom{D}{m}\,$, thus giving the number of degrees of freedom $\Dof_{U(1)\times U(1)}=(D-2)^2$ that corresponds to the transverse polarizations of the tensor $\omega_{\mu|\nu}\,$. The above partition function can be decomposed into its irreducible spacetime components:
\begin{equation}\label{U1U1anatomy}
Z_{U(1)\times U(1)}=t_{1,1}-4\,t_{1,0}+4\,t_{0,0}=[t_{1,1}-t_{2,0}-2\,t_{1,0}]+[t_{2,0}-2\,t_{1,0}+3\,t_{0,0}]+t_{0,0}= Z_h+Z_B+Z_\phi  
\end{equation}
namely the spin two graviton, the two-form and a scalar. The corresponding degrees of freedom decompose accordingly as
\begin{equation}
\Dof_{U(1)\times U(1)}=(D-2)^2=\frac{D(D-3)}{2}+\frac{(D-2)(D-3)}{2}+1    
\end{equation}
and coincide with the transverse polarizations of the (traceless) graviton $h_{ij}\,$, two-form $B_{ij}$ and scalar $\phi\,$.
In the following we will be interested in gauging larger subgroups of $SO(4)$ in order to project out the two-form and/or the scalar field from the spectrum.
In particular we will now analyze how the maximal gauging of the entire $R$-symmetry group, that was studied in \cite{Bastianelli:2007pv} for general $\N\,$, extracts the graviton degrees of freedom.
When the entire $SO(4)$ group is gauged there is no room for Chern-Simons couplings. This fixes the Young diagram of the spacetime gauge field to have $\tfrac{D-2}{2}$ rows, thus yielding a graviton only in $D=4\,$. Restricting to four dimensions, the partition function is given by (see \cite{Bastianelli:2007pv} for the derivation)
\begin{equation}
\begin{split}
Z_{SO(4)}(T) &= \frac{1}{4}\int_0^{2\pi}\frac{d\theta}{2\pi}\int_0^{2\pi}\frac{d\phi}{2\pi}\,\left(2\cos\tfrac{\theta}{2}\right)^{-2}\left(2\cos\tfrac{\phi}{2}\right)^{-2}\left(2i\,\sin\tfrac{\theta+\phi}{2}\right)^2\left(2i\,\sin\tfrac{\theta-\phi}{2}\right)^2\\
&\times\Tr\left[e^{-T\hat H}e^{i\theta(\hat N_1-2)+i\phi(\hat N_2-2)}\right]  \;.  
\end{split}
\end{equation}
In the above expression the cosine factors, as in the previous case, come from the path integral over the SUSY ghosts. The sine factors result instead from the path integral of the non-abelian ghosts of $SO(4)\,$. In particular, the factor containing the difference of the angles corresponds to the $U(2)$ subgroup generated by $J_i^j:=\psi_i\cdot\bar\psi^j\,$, while the other factor, depending on the sum of the angles, is related to the gauging of $\bar K^{ij}:=\bar\psi^i\cdot\bar\psi^j$ (trace operator) and $K_{ij}:=\psi_i\cdot\psi_j$ (insertion of the metric). In terms of Wilson line variables we have
\begin{equation}\label{Zso4}
Z_{SO(4)}(T)=\frac14\oint_{\gamma^-}\frac{dz}{2\pi iz}\oint_{\gamma^-}\frac{dw}{2\pi iw}\,\frac{z}{(z+1)^2}\frac{w}{(w+1)^2}\,p(z,w)\sum_{n,m=0}^Dt_{n,m}(T)\,z^{n-2}w^{m-2} \;, 
\end{equation}
where the function
\begin{equation}\label{so4projector}
p(z,w)=\frac{(zw-1)^2(z-w)^2}{z^2w^2} = 4-2zw-2\left(\frac{z}{w}+\frac{w}{z}\right)-\frac{2}{zw}+z^2+w^2+\frac{1}{z^2}+\frac{1}{w^2}   
\end{equation}
contains all the dependence on the non-abelian gauging. We shall now compute the contributions to \eqref{Zso4} of the various components of $p(z,w)$ in order to see how the projection on the graviton is achieved
\begin{equation}\label{anatomy}
\begin{split}
4\;&\rightarrow\quad t_{1,1}-4\,t_{1,0}+4\,t_{0,0}\equiv Z_{U(1)\times U(1)}  \\[2mm]
-2zw\;&\rightarrow\quad -\tfrac12\,t_{0,0}\equiv -\tfrac12\, Z_\phi \\[2mm]
-2\left(\frac{z}{w}+\frac{w}{z}\right)\;&\rightarrow\quad -(t_{2,0}-2\,t_{1,0}+3\,t_{0,0})\equiv-Z_B\\[2mm]
-\frac{2}{zw}\;&\rightarrow\quad -\tfrac12\,(t_{2,2}+4\,t_{1,1}+9\,t_{0,0}-4\,t_{2,1}+6\,t_{2,0}-12\,t_{1,0})\equiv-\tfrac12\,Z_{A_{2,2}}\\[2mm]
z^2+w^2\;&\rightarrow\quad 0\\[2mm]
\frac{1}{z^2}+\frac{1}{w^2}\;&\rightarrow\quad\tfrac12(t_{3,1}-2\,t_{2,1}+3\,t_{1,1}-4\,t_{0,1})-\tfrac12(t_{3,0}-2\,t_{2,0}+3\,t_{1,0}-4\,t_{0,0})\equiv\tfrac12\,Z_{A_{3,1}}\;,
\end{split}    
\end{equation}
where the arrows mean that the given monomial yields the corresponding contribution upon integration over the moduli $z$ and $w$.

The first three contributions are clear from what we have discussed so far. One can see, for instance, that the third contribution removes the two-form from the reducible partition function $Z_{U(1)\times U(1)}\,$, while the second contribution removes only ``half'' of the scalar field.
To explain the interpretation of the second half of \eqref{anatomy}, let us introduce $(p,q)$ gauge bi-forms. By a gauge bi-form $A_{p,q}$ we denote the tensor field
\begin{equation}
A_{p,q}(x,\psi_i)=A_{\mu[p]\vert \nu[q]}(x)\,\psi_1^{\mu_1}...\psi_1^{\mu_p}\,\psi_2^{\nu_1}...\psi_2^{\nu_q}\sim p\left\{\,\yng(1,1,1,1)\right.\,\otimes\, q\left\{\yng(1,1,1)\,\right.\;,
\end{equation}
with gauge symmetries
\begin{equation}
\delta A_{p,q}=d_1\lambda_{p-1,q}+d_2\xi_{p,q-1}    \;,\quad d_i:=\psi_i^\mu\de_\mu
\end{equation}
and gauge invariant curvature $F_{p+1,q+1}=d_1d_2A_{p,q}\,$. The corresponding partition function can be obtained from the previous case of $U(1)\times U(1)$ gauging by modifying the Chern-Simons couplings to $q_1=p+1-\tfrac{D}{2}$ and $q_2=q+1-\tfrac{D}{2}\,$, yielding
\begin{equation}
Z_{A_{p,q}}=\sum_{k,l=0}^{p,q}(-)^{k+l}(k+1)(l+1)t_{p-k,q-l}\;,    
\end{equation}
that justifies the identifications made in \eqref{anatomy}. 
Let us now consider the contribution of $Z_{A_{2,2}}$ to $Z_{SO(4)}$. By performing a double Hodge dualization of the field $A_{2,2}$ we can see that it is dual to a scalar (recalling that we are in four dimensions):
\begin{equation}
A_{2,2}\;\xrightarrow{\text{curvature}}\;F_{3,3}\;\xrightarrow{\text{Hodge dual}}\;\widetilde F_{1,1}\;\xrightarrow{\text{potential}}\;\widetilde A_{0,0}\equiv\widetilde\phi    \;.
\end{equation}
Similarly, one can see that $A_{3,1}$ is a non-propagating field
\begin{equation}
A_{3,1}\;\xrightarrow{\text{curvature}}\;F_{4,2}\;\xrightarrow{\text{Hodge dual}}\;\widetilde F_{0,2}\;\xrightarrow{\text{potential}}\;\emptyset \end{equation}
with zero degrees of freedom.  We can now put together the results of \eqref{anatomy} and, using the decomposition \eqref{U1U1anatomy}, obtain
\begin{equation}
Z_{SO(4)}=Z_h+\tfrac12\,(Z_\phi-Z_{\widetilde\phi})+\tfrac{1}{2}\,Z_{A_{3,1}}  \;.  
\end{equation}
We can thus conclude that the effective action generated by the full $SO(4)$ gauging corresponds to the graviton plus topological terms. The latter vanish in the free theory, corresponding to zero degrees of freedom, but do contribute to the effective action on non-trivial backgrounds. 
Our goal is to find a modified one-loop measure for the path integral that projects exactly onto the graviton state.
In the measure \eqref{Zso4} the factors of $\frac{z}{(z+1)^2}\frac{w}{(w+1)^2}$ and $p(z,w)$ play very different roles: as we have previously mentioned, the former corresponds to the gauging of worldline supersymmetries, and is responsible of the spacetime gauge symmetry that ensures unitarity. The latter factor, related to the gauging of the $R$-symmetries, performs algebraic projections on the spectrum and is the only one that we will modify.
To begin with, we notice that the $SO(4)$ projector \eqref{so4projector} has the manifest symmetry $p(z,w)=p(1/z, 1/w)$ and can be written as
\begin{equation}
p(z,w)=P(z,w)+P(1/z,1/w)\;,\quad P(z,w)=2-\left(\frac{z}{w}+\frac{w}{z}\right)-2zw+z^2+w^2   \;. 
\end{equation}
By looking at the decomposition \eqref{anatomy}, it is clear that the unwanted contributions from dual fields come from the term $P(1/z,1/w)\,$, that indeed entails double Hodge dualization. We will thus propose to use $2P(z,w)$ instead of $p(z,w)$ as an ansatz for the graviton projector:  
\begin{equation}\label{Zgrav}
Z_{\rm grav}(T)=\frac12\oint_{\gamma^-}\frac{dz}{2\pi iz}\oint_{\gamma^-}\frac{dw}{2\pi iw}\,\frac{z}{(z+1)^2}\frac{w}{(w+1)^2}\,P(z,w)\Tr\left[e^{-T\hat H}z^{\hat N_1-2}w^{\hat N_2-2}\right] \;. 
\end{equation}
This ansatz for the projector can be related to a different gauging of the $R$-symmetry algebra.
First, notice that $P(z,w)$ can be written in factorized form as
\begin{equation}
P(z,w)=\frac{(zw-1)(z-w)^2}{zw}\;,\;{\rm or}\quad P(\theta,\phi)=2i\,\sin\tfrac{\theta+\phi}{2}\left(2i\,\sin\tfrac{\theta-\phi}{2}\right)^2\exp{i\left(\tfrac{\theta+\phi}{2}\right)}\;. \end{equation}
We now consider the case of gauging the parabolic subalgebra of $so(4)$ consisting of the $u(2)$ subalgebra, generated by $J_i^j\,$, plus the trace $\bar K^{ij}\,$. We are thus excluding, compared to the maximal gauging, the insertion of the metric $K_{ij}\,$, which causes in the measure the depletion of one sine factor, yielding 
\begin{align}
P_{\rm parab}(\theta,\phi)&=2i\,\sin\tfrac{\theta+\phi}{2}\left(2i\,\sin\tfrac{\theta-\phi}{2}\right)^2\;,
\label{eq:parabolic}
\\
%\;\text{that is}\quad 
\Rightarrow\ P(\theta,\phi)&= \exp{i\left(\tfrac{\theta+\phi}{2}\right)}P_{\rm parab}(\theta,\phi)  \;.
\end{align}
The extra exponential factor can be accounted for, since the parabolic gauging allows for a Chern-Simons term proportional to the diagonal $U(1)\,$:
\begin{equation}
S_{\rm CS}=iq\,\int d\tau\,a^i_i\quad\stackrel{\text{gauge fix}}{\longrightarrow}\quad iq(\theta+\phi)\;.  \end{equation}
By an appropriate choice of $q$ it is possible to reproduce the graviton projector $P(\theta,\phi)$ and also to go in arbitrary dimensions. In fact, recall that the shift in the number operators in \eqref{Zso4} is in general $\hat N_i-\tfrac{D}{2}\,$. By choosing the Chern-Simons coupling as
$$
q=-\frac{D-3}{2}=2-\frac{D}{2}-\frac12\;,
$$
the $\tfrac12$ part produces $P(\theta,\phi)$ from $P_{\rm parab}(\theta,\phi)\,$, while the rest modifies the shifts $\hat N_i-\tfrac{D}{2}$ to $\hat N_i-2$ in arbitrary dimensions.

\section{One loop gravity effective action}
In the present section we apply the method described above to represent the one-loop effective action for pure Einstein-Hilbert gravity, 
and compute it in a local expansion up to quadratic orders in the 
curvature\footnote{For massless particles the local expansion of the one-loop effective action is not permitted, except for identifying the
divergences which indeed are local. Here we consider precisely those terms.}. 
The starting point is the $SO(4)$-extended locally supersymmetric spinning particle, whose phase-space action corresponds to the curved deformation of action~\eqref{eq:4flat}. It was previously studied in~\cite{Bastianelli:2012bn}, where a suitable BRST gauge fixing, involving the whole $R$-symmetry group, was analyzed. 

In the present approach, we instead consider the gauge fixing of the parabolic subgroup discussed above, which corresponds to the phase space action
\begin{align}
	S[z,E;g]=\int dt \Big[p_\mu\dot x^\mu+i\,\bar\psi^{i\,\mu}\dot\psi_{i\,\mu}-e\,H-i\chi_i\,\bar\psi^i\cdot \pi-i\bar\chi^i\,\psi_i\cdot \pi-\frac12 a_{ij}\bar K^{ij}-a_i{}^j(J_j^i-iq\delta_j^i)\Big]\,,
\end{align}
where $z=(x,p,\psi,\bar\psi)$, $E=(e,\chi_i,\bar\chi^i,a_{ij},a_i{}^j)$, and
\begin{align}
	\pi_\mu &= p_\mu-\omega_{\mu ab}\,\psi_i^a\bar\psi^{ib}\\
	H&=  g^{\mu\nu}\pi_\mu \pi_\nu- R_{abcd}\, \psi_i^a \bar\psi^{ib} \psi_j^c\bar\psi^{jd}~.
\end{align}
The one-loop effective action thus corresponds to the circle path integral of the previous action. The antiperiodic gravitini $\chi_i,\, \bar\chi^i$ are again gauge-fixed to zero, whereas the einbein gets gauge-fixed to its modulus, $T$, which is interpreted as the Schwinger proper-time, and the gauge fields of the parabolic subgroup are fixed to the two angles of the associated Cartan torus, with the Faddeev-Popov determinant given by the expression~\eqref{eq:parabolic}. In euclidean configuration space we thus have
\begin{equation}\label{3.4}
\Gamma[g]= -\frac12 \int_0^\infty\frac{dT}{T}Z(T)\;,    
\end{equation}
with
\begin{equation} 
\begin{split}
Z(T)&=\frac12 \int_0^{2\pi}\frac{d\theta}{2\pi} \int_0^{2\pi}\frac{d\phi}{2\pi}\left(2\cos \frac{\theta}{2}\right)^{-2} \left(2\cos \frac{\phi}{2}\right)^{-2}\, P_{\rm parab}(\theta,\phi)\, e^{-iq(\theta+\phi)}\\
	&\times \int_P
	{\cal D}x \int_A D\bar \psi D\psi \exp\Big\{-\int_0^1 d\tau\Big[\,\frac1{4T} \,g_{\mu\nu}\dot x^\mu \dot x^\nu +\bar\psi^{ia}(D_\tau\delta_i^j-\Lambda_i{}^j)\psi_{ja}\\
	&\hspace{5cm}-T R_{abcd}\,\psi^a\cdot \bar\psi^b \psi^c\cdot \bar\psi^d +2T\,V_{\rm im}\Big]\Big\}\;,
\end{split}
\end{equation}
where 
$
	\Lambda_i{}^j=\scaleobj{0.75}{\left(\begin{array}{cc}
		\theta &0\\
		0&\phi
	\end{array}\right)}
$, $D_\tau$ is the covariant derivative in the multispinor representation of the Lorentz group,
and the scalar potential $V_{\rm im}=-\tfrac{D+2}{8(D-1)}R$ is an order $\hbar^2$ improvement term, generated at the quantum level from the anticommutator of the supersymmetry generators. 
The constraints are not first class, but the expression \eqref{3.4} coincides with the one coming from the BRST system of ref.~\cite{Bonezzi:2018box}, considered as the starting point for the path integral. The BRST system, and thus the path integral, is only consistent (upon projection) on Einstein manifolds: $R_{\mu\nu}=\lambda\,g_{\mu\nu}$, which is the class of allowed backgrounds
\cite{Bonezzi:2018box}.

The above curved space particle path integral involves a nonlinear sigma model action, whose perturbative (short time) evaluation 
needs a careful regularization. This is well studied (see ref.~\cite{Bastianelli:2006rx} for a review), even in models with extended supersymmetries~\cite{Bastianelli:2011cc}. In particular, the use of worldline dimensional regularization implies the inclusion of a counterterm potential which, for $\N=4$, is
$
	V_{\rm CT}=\frac18 R
$
and, together with the improvement term gives
\begin{align}
	V=\left(\frac18 -\frac{D+2}{8(D-1)}\right)R = -\frac{3}{8(D-1)}R =: \omega R\,.
\end{align}
With these prescriptions, and with the Feynman rules described in  \cite{Bastianelli:2012bn}, we obtain
\begin{align}
	Z(T)&=\frac12 \int_0^{2\pi}\frac{d\theta}{2\pi} \int_0^{2\pi}\frac{d\phi}{2\pi}\left(2\cos \frac{\theta}{2}\right)^{D-2} \left(2\cos \frac{\phi}{2}\right)^{D-2}\, P_{\rm parab}(\theta,\phi)\, e^{-iq(\theta+\phi)}\nonumber\\
	&\times\int d^Dx\, \frac{\sqrt{g}} {(4\pi T)^{\frac D2}}\Big\langle e^{-S_{int}}\Big\rangle\nonumber\\&=:
	\int d^Dx\, \frac{\sqrt g}{(4\pi T)^{\frac D2}}\,
	\Big\langle\!\!\Big\langle e^{-S_{int}}\Big\rangle\!\!\Big\rangle
	\,,
	\label{3.7}
\end{align}
where
\begin{align}
    \Big\langle e^{-S_{int}}\Big\rangle &= 1-T\Big(-\tfrac{5}{12}+2\omega +\tfrac14 \big( \cos^{-2}\tfrac{\theta}{2}+\cos^{-2}\tfrac{\phi}{2}\big)\Big)R\nonumber\\
    &+T^2\Big\{\tfrac12  \Big(-\tfrac{5}{12}+2\omega +\tfrac14\big( \cos^{-2}\tfrac{\theta}{2}+\cos^{-2}\tfrac{\phi}{2}\big)\Big)^2R^2\nonumber\\
    &+\Big( -\tfrac1{180}-\tfrac1{8}\big( \cos^{-4}\tfrac{\theta}{2}+\cos^{-4}\tfrac{\phi}{2}\big)+\tfrac1{8}\big( \cos^{-2}\tfrac{\theta}{2}+\cos^{-2}\tfrac{\phi}{2}\big)\Big)R_{ab}^2\nonumber\\
    &+\Big( \tfrac1{180}-\tfrac1{32}\big( \cos^{-4}\tfrac{\theta}{2}+\cos^{-4}\tfrac{\phi}{2}\big)-\tfrac1{48}\big( \cos^{-2}\tfrac{\theta}{2}+\cos^{-2}\tfrac{\phi}{2}\big)+\tfrac1{16}\big( \cos^{-2}\tfrac{\theta}{2}+\cos^{-2}\tfrac{\phi}{2}\big)^2\Big)R_{abcd}^2\nonumber\\
    &-\Big( -\tfrac{9}{120}+\tfrac{\omega}{3}+\tfrac{1}{24} \big( \cos^{-2}\tfrac{\theta}{2}+\cos^{-2}\tfrac{\phi}{2}\big)\Big)\nabla^2R+O(T^3)\Big\}~,
\end{align}
and $\langle\langle \dots \rangle \rangle$ is the modular integral of the path integral average. 
As above we find it convenient to rewrite the modular integrals in terms of complex variables rather than angles. We thus have
\begin{align}
    \Big\langle\!\!\Big\langle e^{-S_{int}}\Big\rangle\!\!\Big\rangle &= \frac12 \oint \frac{dz}{2\pi i z} \oint \frac{dw}{2\pi i w} \frac{(1+z)^{D-2}}{z^2}\frac{(1+w)^{D-2}}{w^2} (zw-1)(z-w)^2 \nonumber
    \\
    & \Biggl\{1-T\Big(-\tfrac{5}{12}+2\omega + \tfrac{z}{(1+z)^2}+\tfrac{w}{(1+w)^2}\Big)R\nonumber\\
    &+T^2\Big[\tfrac12  \Big(-\tfrac{5}{12}+2\omega + \tfrac{z}{(1+z)^2}+\tfrac{w}{(1+w)^2}\Big)^2R^2\nonumber\\
    &+\Big( -\tfrac1{180}-2\big( \tfrac{z^2}{(1+z)^4}+\tfrac{w^2}{(1+w)^4}\big)+\tfrac12 \big( \tfrac{z}{(1+z)^2}+\tfrac{w}{(1+w)^2}\big)\Big)R_{ab}^2\nonumber\\
    &+\Big( \tfrac1{180}-\tfrac1{2}\big( \tfrac{z^2}{(1+z)^4}+\tfrac{w^2}{(1+w)^4}\big)-\tfrac1{12}\big( \tfrac{z}{(1+z)^2}+\tfrac{w}{(1+w)^2}\big)+\big( \tfrac{z}{(1+z)^2}+\tfrac{w}{(1+w)^2}\big)^2\Big)R_{abcd}^2\nonumber\\
    &-\Big( -\tfrac{9}{120}+\tfrac{\omega}{3}+\tfrac{1}{6} \big( \tfrac{z}{(1+z)^2}+\tfrac{w}{(1+w)^2}\big)\Big)\nabla^2R+O(T^3)\Big]\Biggr\}\;,
\end{align}
which, after performing the contour integrals, yields
\begin{align}
     \Big\langle\!\!\Big\langle e^{-S_{int}}\Big\rangle\!\!\Big\rangle &=\frac{D(D-3)}{2}\Biggl\{1 +T R\, \frac{5D^2-35D+12}{12(D-1)(D-3)}\nonumber\\
     &+T^2 \Biggl[R^2\, \frac{25 D^4 - 275 D^3 + 232 D^2- 432 D+288  }{288  D(D-1)^2 (D-3)}%\nonumber\\&
     - R_{ab}^2\, \frac{ D^2- 183 D-720}{180 D(D-3)}\nonumber\\
     & + R_{abcd}^2\, \frac{D^2- 33 D+540}{180 D (D-3)}+ \nabla^2 R\, \frac{ 9 D^2- 61 D+22}{120 ( D-1) (D-3)}\Biggr]%\nonumber\\&
     +O(T^3)\Biggr\} \;.
\end{align}

For $D=4$ it reduces to 
\begin{align}
     \Big\langle\!\!\Big\langle e^{-S_{int}}\Big\rangle\!\!\Big\rangle &=\Biggl\{2 -  \frac83\,T\, R  
     +T^2 \Biggl[ -\frac{31}{18} R^2 + \frac{359}{90}
     R_{ab}^2 
      +\frac{53}{45} R_{abcd}^2 -\frac{13}{30} \nabla^2 R \Biggr]%\nonumber\\&
     +O(T^3)\Biggr\} \;,
\end{align}
that on
Einstein manifolds, coincides with the expression found in ref.~\cite{Bastianelli:2013tsa}, namely
\begin{align}
     \Big\langle\!\!\Big\langle e^{-S_{int}}\Big\rangle\!\!\Big\rangle &=\Biggl\{2 -   \frac83\,T\, R  
     +T^2 \Biggl[ -\frac{29}{40} R^2 
      +\frac{53}{45} R_{abcd}^2 \Biggr]%\nonumber\\&
     +O(T^3)\Biggr\} \;,
     \label{eq:Einstein-onshell}
\end{align}
where $R_{abcd}^2$ can be identified with the four-dimensional Euler density
of Einstein manifolds.
The different powers of $T$, once inserted into \eqref{3.7} and \eqref{3.4},  give rise to the quartic, quadratic and logarithmic divergences 
of the graviton one-loop effective action.
In spacetime dimensional regularization the first two terms are invisible, and the 
logarithmic divergence vanishes for null cosmological constant, thus reproducing the famous result by 't Hooft and Veltman \cite{tHooft:1974toh}.  
With a cosmological constant, the logarithmic divergence coincides with the one found in \cite{Christensen:1979iy}.
More generally, all these divergences reproduce those computed in \cite{Bastianelli:2013tsa}.

The on-shell expression given in eq.~\eqref{eq:Einstein-onshell}  is gauge independent, and it is a benchmark for any correct calculation in
perturbative quantum gravity. Thus, for completeness, we also report the same result for a graviton on a $D$-dimensional Einstein manifold
\begin{align}
     \Big\langle\!\!\Big\langle e^{-S_{int}}\Big\rangle\!\!\Big\rangle &=\frac{D(D-3)}{2} +T R\, \frac{D(5D^2-35D+12)}{24(D-1)}\nonumber\\
     &+T^2 \Biggl[R^2\, \frac{125 D^5-1383D^4+2640D^3+664D^2-8616D+5760}{2880 D(D-1)^2}%\nonumber\\&
     \nonumber\\
     & + R_{abcd}^2\, \frac{D^2- 33 D+540}{360}
     %+ \nabla^2 R\, \frac{ 9 D^2- 61 D+22}{960 D( D-1)}
     \Biggr]%\nonumber\\&
     +O(T^3)\;,
\end{align}
where the first term gives the number of its degrees of freedom.

%However, the above result is expected to hold in any space-time dimension.  For instance it is intriguing to consider the case $D=3$, for which Einstein-Hilbert gravity has no propagating degrees of freedom but there are UV linear divergences at one loop, which correspond to the term linear in $R$. 

\section{Conclusions}
We have constructed a modified  $\mathcal{N}=4$  spinning particle that is able to describe the pure graviton on Einstein spaces.
It is identified by gauging only a parabolic subgroup of the $SO(4)$ $R$-symmetry group of the particle, 
and adding suitable Chern-Simons couplings on the worldline.
The gauging of parabolic subgroups have already been used in worldline models for higher spin particles, as in ref.~\cite{Bastianelli:2009eh}. They give rise to worldline actions that are not real, but which do not seem to produce
any pathology in the spacetime intepretation of the theory, in that the sole purpose of the $R$-symmetry gauging is to perform an algebraic projection on the spacetime spectrum to achieve different degrees of reducibility.
Similarly, the use of Chern-Simons couplings on the worldline helps to insert projectors in the Hilbert space, so to leave
only a desired subsector containing the physical states. They appeared, for example,  
in the worldline description of differential forms \cite{Bastianelli:2005vk, Bastianelli:2011pe, Bastianelli:2012nh}.
 
 Our model has been able to reproduce the correct divergences of one-loop quantum gravity.
 Together with the BRST construction of ref. \cite{Bonezzi:2018box}, few working tools are now available to address quantum gravity from a worldline perspective.
 
We stress again that the model developed in this paper can be naively obtained by gauging the worldline supersymmetries, together with appropriate subalgebras of the $R$-symmetry. However, the classical supersymmetry algebra is broken by the background curvature, and the action has to be seen as a quantum BRST model, whose consistency is guaranteed by the truncation of the Hilbert space. On the reduced Hilbert space the BRST charge is nilpotent only when the background is on-shell according to Einstein's equation \cite{Bonezzi:2018box}, showing that the ``prediction'' of target space equations from quantum consistency of the first-quantized theory is not a peculiarity of string theory, as already confirmed by the earlier work of ref. \cite{Dai:2008bh}.   

It would be useful to extend further these constructions and find more applications of worldline methods to quantum gravity, for instance extending the present description to non-commutative spaces \cite{Bonezzi:2012vr}.
An immediate generalization of the present model consists in weakening the $R$-symmetry constraint to $U(1)\times U(1)\,$, thereby letting all the massless NS-NS particles circulate in the loop. Further coupling the model to a background $B$-field and dilaton \cite{Bonezzi:new} would allow to obtain the one-loop effective action for the whole NS-NS massless sector of string theory.

An alternative to the present formulation in terms of fermionic oscillators is to use the 
$Sp(2)$ particle, with bosonic oscillators. The advantage, once a consistent BRST system is found, would be to easily treat all spins at once, by just modifying a suitable Chern-Simons coupling. This should also reproduce the well known no-go results for minimal coupling higher spin fields to gravity \cite{Aragone:1979hx,Fradkin:1987ks}.


\begin{thebibliography}{99}

%\cite{Bonezzi:2018box}
\bibitem{Bonezzi:2018box}
  R.~Bonezzi, A.~Meyer and I.~Sachs,
  ``Einstein gravity from the $ \mathcal{N}=4 $ spinning particle,''
  JHEP {\bf 1810} (2018) 025
    [\href{https://arxiv.org/abs/1807.07989}{\tt arXiv:1807.07989}].
%  doi:10.1007/JHEP10(2018)025
   %%CITATION = doi:10.1007/JHEP10(2018)025;%%

%\cite{Berezin:1976eg}
\bibitem{Berezin:1976eg}
  F.~A.~Berezin and M.~S.~Marinov,
  ``Particle spin dynamics as the Grassmann variant of classical mechanics,''
  Annals Phys.\  {\bf 104} (1977) 336.
 % doi:10.1016/0003-4916(77)90335-9
  %%CITATION = doi:10.1016/0003-4916(77)90335-9;%%
  %635 citations counted in INSPIRE as of 01 Aug 2019
 
%\cite{Gershun:1979fb}
\bibitem{Gershun:1979fb}
  V.~D.~Gershun and V.~I.~Tkach,
  ``Classical and quantum dynamics of particles with arbitrary spin,''
  JETP Lett.\  {\bf 29} (1979) 288
   [Pisma Zh.\ Eksp.\ Teor.\ Fiz.\  {\bf 29} (1979) 320].
  %%CITATION = JTPLA,29,288;%%
  %142 citations counted in INSPIRE as of 01 Aug 2019

%\cite{Howe:1988ft}
\bibitem{Howe:1988ft}
  P.~S.~Howe, S.~Penati, M.~Pernici and P.~K.~Townsend,
  ``Wave equations for arbitrary spin from quantization of the 
extended supersymmetric spinning particle,''
  Phys.\ Lett.\ B {\bf 215} (1988) 555.
   %%CITATION = doi:10.1016/0370-2693(88)91358-5;%%
  %148 citations counted in INSPIRE as of 01 Aug 2019

%\cite{Siegel:1988ru}
\bibitem{Siegel:1988ru}
  W.~Siegel,
  ``Conformal invariance of extended spinning particle mechanics,''
  Int.\ J.\ Mod.\ Phys.\ A {\bf 3} (1988) 2713.
 % doi:10.1142/S0217751X88001132
  %%CITATION = doi:10.1142/S0217751X88001132;%%
  %70 citations counted in INSPIRE as of 01 Aug 2019

%\cite{Siegel:1988gd}
\bibitem{Siegel:1988gd}
  W.~Siegel,
  ``All free conformal representations in all dimensions,''
  Int.\ J.\ Mod.\ Phys.\ A {\bf 4} (1989) 2015.
  %doi:10.1142/S0217751X89000819
  %%CITATION = doi:10.1142/S0217751X89000819;%%
  %91 citations counted in INSPIRE as of 01 Aug 2019

%\cite{Kuzenko:1995mg}
\bibitem{Kuzenko:1995mg}
  S.~M.~Kuzenko and Z.~V.~Yarevskaya,
  ``Conformal invariance, N extended supersymmetry and massless spinning particles in anti-de Sitter space,''
  Mod.\ Phys.\ Lett.\ A {\bf 11} (1996) 1653
 % doi:10.1142/S0217732396001648
   [\href{https://arxiv.org/abs/hep-th/9512115}{{\tt hep-th/9512115}}].
  %%CITATION = doi:10.1142/S0217732396001648;%%
  %31 citations counted in INSPIRE as of 01 Aug 2019

%\cite{Bastianelli:2008nm}
\bibitem{Bastianelli:2008nm}
  F.~Bastianelli, O.~Corradini and E.~Latini,
  ``Spinning particles and higher spin fields on (A)dS backgrounds,''
  JHEP {\bf 0811} (2008) 054
%  doi:10.1088/1126-6708/2008/11/054
   [\href{https://arxiv.org/abs/0810.0188}{{\tt arXiv:0810.0188}}].
  %%CITATION = doi:10.1088/1126-6708/2008/11/054;%%
  
 %\cite{Corradini:2010ia}
\bibitem{Corradini:2010ia}
  O.~Corradini,
  ``Half-integer higher spin fields in (A)dS from spinning particle models,''
  JHEP {\bf 1009} (2010) 113
   [\href{ https://arxiv.org/abs/1006.4452}{{\tt arXiv:1006.4452}}].
 %  doi:10.1007/JHEP09(2010)113
   %%CITATION = doi:10.1007/JHEP09(2010)113;%% 

  %\cite{Bastianelli:2013tsa}
\bibitem{Bastianelli:2013tsa}
  F.~Bastianelli and R.~Bonezzi,
  ``One-loop quantum gravity from a worldline viewpoint,''
  JHEP {\bf 1307} (2013) 016
   [\href{https://arxiv.org/abs/1304.7135}{{\tt arXiv:1304.7135}}].
%  doi:10.1007/JHEP07(2013)016
   %%CITATION = doi:10.1007/JHEP07(2013)016;%%
  %15 citations counted in INSPIRE as of 31 Jul 2019
  
\bibitem{Dai:2008bh}
  P.~Dai, Y.~t.~Huang and W.~Siegel,
  ``Worldgraph approach to Yang-Mills amplitudes from N=2 spinning particle,''
  JHEP {\bf 0810} (2008) 027
    [\href{https://arxiv.org/abs/0807.0391}{\tt
   arXiv:0807.0391}].
    %%CITATION = doi:10.1088/1126-6708/2008/10/027;%%

\bibitem{Bekaert:2004dz}
  X.~Bekaert, N.~Boulanger and S.~Cnockaert,
  ``No self-interaction for two-column massless fields,''
  J.\ Math.\ Phys.\  {\bf 46} (2005) 012303
 % doi:10.1063/1.1823032
  [\href{https://arxiv.org/abs/hep-th/0407102}{\tt hep-th/0407102}].
  
\bibitem{Bastianelli:2007pv}
  F.~Bastianelli, O.~Corradini and E.~Latini,
  ``Higher spin fields from a worldline perspective,''
  JHEP {\bf 0702} (2007) 072
  [\href{https://arxiv.org/abs/hep-th/0701055}{{\tt hep-th/0701055}}].
%doi:10.1088/1126-6708/2007/02/072
   %%CITATION = doi:10.1088/1126-6708/2007/02/072;%%

%\cite{tHooft:1974toh}
\bibitem{tHooft:1974toh}
  G.~'t Hooft and M.~J.~G.~Veltman,
  ``One loop divergencies in the theory of gravitation,''
  Ann.\ Inst.\ H.\ Poincare Phys.\ Theor.\ A {\bf 20} (1974) 69.
  %976 citations counted in INSPIRE as of 01 Aug 2019
 
 %\cite{Christensen:1979iy}
\bibitem{Christensen:1979iy}
  S.~M.~Christensen and M.~J.~Duff,
  ``Quantizing gravity with a cosmological constant,''
  Nucl.\ Phys.\ B {\bf 170} (1980) 480.
 % doi:10.1016/0550-3213(80)90423-X
  %%CITATION = doi:10.1016/0550-3213(80)90423-X;%%
  %261 citations counted in INSPIRE as of 01 Aug 2019
 
\bibitem{Howe:1989vn}
  P.~S.~Howe, S.~Penati, M.~Pernici and P.~K.~Townsend,
  ``A particle mechanics description of antisymmetric tensor fields,''
  Class.\ Quant.\ Grav.\  {\bf 6} (1989) 1125.
    
\bibitem{Bastianelli:2005vk}
  F.~Bastianelli, P.~Benincasa and S.~Giombi,
  ``Worldline approach to vector and antisymmetric tensor fields,''
  JHEP {\bf 0504} (2005) 010
  [\href{https://arxiv.org/abs/hep-th/0503155}{{\tt
  hep-th/0503155}}].
  
  %\cite{Siegel:1989ip}
\bibitem{Siegel:1989ip}
  W.~Siegel,
  ``Relation between Batalin-Vilkovisky and first quantized style BRST,''
  Int.\ J.\ Mod.\ Phys.\ A {\bf 4} (1989) 3705.
%  doi:10.1142/S0217751X89001485
  %%CITATION = doi:10.1142/S0217751X89001485;%%
  
\bibitem{Barnich:2003wj}
  G.~Barnich and M.~Grigoriev,
  ``Hamiltonian BRST and Batalin-Vilkovisky formalisms for second quantization of gauge theories,''
  Commun.\ Math.\ Phys.\  {\bf 254} (2005) 581
  [\href{https://arxiv.org/abs/hep-th/0310083}{{\tt
  hep-th/0310083}}].
  
\bibitem{Barnich:2004cr}
  G.~Barnich, M.~Grigoriev, A.~Semikhatov and I.~Tipunin,
  ``Parent field theory and unfolding in BRST first-quantized terms,''
  Commun.\ Math.\ Phys.\  {\bf 260} (2005) 147
 [\href{https://arxiv.org/abs/hep-th/0406192}{{\tt
  hep-th/0406192}}].
  
\bibitem{Barnich:2006pc}
  G.~Barnich and M.~Grigoriev,
  ``Parent form for higher spin fields on anti-de Sitter space,''
  JHEP {\bf 0608} (2006) 013
 [\href{https://arxiv.org/abs/hep-th/0602166}{{\tt
  hep-th/0602166}}].
  
  %\cite{Bastianelli:2012bn}
\bibitem{Bastianelli:2012bn}
  F.~Bastianelli, R.~Bonezzi, O.~Corradini and E.~Latini,
  ``Effective action for higher spin fields on (A)dS backgrounds,''
  JHEP {\bf 1212} (2012) 113
      [\href{https://arxiv.org/abs/1210.4649}{\tt arXiv:1210.4649}].
  %doi:10.1007/JHEP12(2012)113
%  [arXiv:1210.4649 [hep-th]].
  %%CITATION = doi:10.1007/JHEP12(2012)113;%%
   
  %\cite{Bastianelli:2006rx}
\bibitem{Bastianelli:2006rx}
  F.~Bastianelli and P.~van Nieuwenhuizen,
  ``Path integrals and anomalies in curved space,'' Cambridge University Press (Cambridge UK, 2006).
  %doi:10.1017/CBO9780511535031
  %%CITATION = doi:10.1017/CBO9780511535031;%%
 
  %\cite{Bastianelli:2011cc}
\bibitem{Bastianelli:2011cc}
  F.~Bastianelli, R.~Bonezzi, O.~Corradini and E.~Latini,
  ``Extended SUSY quantum mechanics: transition amplitudes and path integrals,''
  JHEP {\bf 1106} (2011) 023
    [\href{https://arxiv.org/abs/1103.3993}{\tt arXiv:1103.3993}].
  %doi:10.1007/JHEP06(2011)023
%  [arXiv:1103.3993 [hep-th]].
  %%CITATION = doi:10.1007/JHEP06(2011)023;%%
 
  
%\cite{Bastianelli:2009eh}
\bibitem{Bastianelli:2009eh}
  F.~Bastianelli, O.~Corradini and A.~Waldron,
  ``Detours and paths: BRST complexes and worldline formalism,''
  JHEP {\bf 0905} (2009) 017
   [\href{https://arxiv.org/abs/0902.0530}{\tt arXiv:0902.0530}].
 % [arXiv:0902.0530 [hep-th]].
  %%CITATION = doi:10.1088/1126-6708/2009/05/017;%%
  %26 citations counted in INSPIRE as of 02 Aug 2019
 
  %\cite{Bastianelli:2011pe}
\bibitem{Bastianelli:2011pe}
  F.~Bastianelli and R.~Bonezzi,
  ``Quantum theory of massless (p,0)-forms,''
  JHEP {\bf 1109} (2011) 018
       [\href{https://arxiv.org/abs/1107.3661}{\tt arXiv:1107.3661}].
    %%CITATION = doi:10.1007/JHEP09(2011)018;%%
  %12 citations counted in INSPIRE as of 02 Aug 2019
  
 %\cite{Bastianelli:2012nh}
\bibitem{Bastianelli:2012nh}
  F.~Bastianelli, R.~Bonezzi and C.~Iazeolla,
  ``Quantum theories of (p,q)-forms,''
  JHEP {\bf 1208} (2012) 045
    [\href{https://arxiv.org/abs/1204.5954}{\tt arXiv:1204.5954}].
  %%CITATION = doi:10.1007/JHEP08(2012)045;%%
  %13 citations counted in INSPIRE as of 02 Aug 2019 
  
  %\cite{Bonezzi:2012vr}
\bibitem{Bonezzi:2012vr}
  R.~Bonezzi, O.~Corradini, S.~A.~Franchino Vi\~nas and P.~A.~G.~Pisani,
  ``Worldline approach to noncommutative field theory,''
  J.\ Phys.\ A {\bf 45} (2012) 405401
 [\href{https://arxiv.org/abs/1204.1013}{\tt arXiv:1204.1013}].
  %%CITATION = doi:10.1088/1751-8113/45/40/405401;%%
  %13 citations counted in INSPIRE as of 02 Aug 2019
 
\bibitem{Bonezzi:new}
  %\cite{Bonezzi:2020jjq}
R.~Bonezzi, A.~Meyer and I.~Sachs,
``A Worldline Theory for Supergravity,''
JHEP \textbf{2006} (2020) 103
[\href{https://arxiv.org/abs/2004.06129}{\tt arXiv:2004.06129}].
%4 citations counted in INSPIRE as of 23 Feb 2021
 
%\cite{Aragone:1979hx}
\bibitem{Aragone:1979hx}
  C.~Aragone and S.~Deser,
  ``Consistency problems of hypergravity,''
  Phys.\ Lett.\  {\bf 86B} (1979) 161.
  %doi:10.1016/0370-2693(79)90808-6
  %%CITATION = doi:10.1016/0370-2693(79)90808-6;%%
  %244 citations counted in INSPIRE as of 20 Aug 2019 
  
%\cite{Fradkin:1987ks}
\bibitem{Fradkin:1987ks}
  E.~S.~Fradkin and M.~A.~Vasiliev,
  ``On the gravitational interaction of massless higher spin fields,''
  Phys.\ Lett.\ B {\bf 189} (1987) 89.
  %doi:10.1016/0370-2693(87)91275-5
  %%CITATION = doi:10.1016/0370-2693(87)91275-5;%%
  %422 citations counted in INSPIRE as of 20 Aug 2019  
 
\end{thebibliography}
\end{document}